\documentclass[aps,pre,twocolumn,superscriptaddress,groupedaddress,floatfix]{revtex4}  
\usepackage{graphicx}   
\usepackage{amssymb}   
\usepackage{amsmath} 
\usepackage{float}

\begin{document}

\widetext

\title{Nonclassical heat flow in passive chiral solids is third rank, not odd}
\author{R. S. Lakes}
\email{lakes@engr.wisc.edu}
\affiliation{Department of Engineering Physics, Department of Materials Science\\University of Wisconsin, Madison, WI 53706}
\date{\today}

\begin{abstract}
	 Chiral, directionally isotropic gyroid lattices are observed to exhibit nonclassical thermal effects incompatible with an asymmetric (``odd'') second rank conductivity tensor but consistent with a third rank tensor property that provides a curl term.  The lattices are passive materials so no driving torques are needed to obtain transverse flow.  A method for determination of the length scale associated with chirality is provided. The polymer gyroid lattice made by 3D printing is a metamaterial which allows chirality that is tunable by geometry. 
\end{abstract}

\maketitle

\section{Introduction} \label{intro}
Symmetry of the even rank tensors that govern dielectric permittivity, elasticity and magnetic permeability follows from the assumption of a conserved energy density \cite{Nye} that pertains to reversible processes between equilibrium states. Conduction of heat or electricity requires non-equilibrium thermodynamics. Tensors that govern coupled processes such as heat and electrical flux obey reciprocity via the Onsager relations \cite{Onsager1, Onsager2} due to time-reversal symmetry. The Onsager relations also require transport coefficient tensors such as thermal conductivity and diffusion to be symmetric but the inference is more indirect \cite{Nye}  \cite{Casimir}. 
\par
   An antisymmetric part of material property tensors has  been called  ``odd''  in the context of the viscosity tensor \cite{Avron1998} and the elastic modulus tensor;  that terminology is used here.   Asymmetric property tensors have long been known. The Hall effect \cite{Hall1879} can be analyzed  \cite{Nye} via an asymmetric second rank tensor. Time reversal symmetry is broken by a magnetic field  \cite{Onsager2} so an asymmetric tensor is possible even though the material itself is usually assumed to be homogeneous and isotropic.  Coupling to a magnetic field in gyrotropic materials gives rise to the Faraday effect that can be incorporated in an asymmetric second rank permittivity tensor \cite{Pershan1967}. The possibility of asymmetry of the thermal conductivity tensor was studied \cite{Soret1893}  \cite{Soret1894} more than 120 years ago. Experimental evidence of asymmetric effects was sought but not found. 
\par
The viscosity tensor and the elasticity tensor are fourth rank. Symmetry with respect to exchange of the first and second pairs of indices is demonstrated via an assumption of a conserved energy density \cite{Nye} \cite{Sokolnikoff}. If the assumption of time reversal symmetry is relaxed, then an asymmetric viscosity is possible \cite{Avron1998}. However ``odd''  effects cannot occur in 3D isotropic viscous or elastic materials but they are possible in 2D isotropy \cite{Avron1995}. Analysis was adduced of possible odd effects in a quantum Hall fluid  \cite{Avron1995} and in superfluid He$^{3}$. Active matter with globally or locally aligned spinning components was shown analytically to allow odd viscosity \cite{Markovich21}.  Chiral active fluids were shown via analysis to exhibit odd diffusive mobility in 2D \cite{Poggioli23}.   Hall-like heat transfer was demonstrated  in an active chiral lattice containing a matrix and a rotating 6 x 6 array of motor driven discs   \cite{Hallheat23}.  
\par
Chirality is of interest in that it entails additional freedom that pertains to passive materials as well as the active materials that have been studied recently. Chirality is essential for physical properties such as piezoelectricity and pyroelectricity. They are described by tensors of odd rank: third rank for piezoelectricity and rank 1 for pyroelectricity  \cite{Nye}. Chirality has no effect on tensors of even rank because an inversion of coordinates    was shown to have   no effect on these tensors. For example, chirality has no effect in classical elasticity because the    symmetric elastic modulus tensor  \cite{Nye}  \cite{Sokolnikoff}      is fourth rank. Elastic chirality nonetheless gives rise to observable effects. Cosserat \cite{Cosserat09}  (micropolar \cite{Eringen68}) elasticity theory provides sufficient freedom to accommodate chiral elastic effects   such as stretch-twist coupling, size effects in torsion and compression, and radial dependence of the Poisson effect.  
More freedom of this sort must be incorporated in elasticity theory to correctly predict phenomena in non-chiral solids with microstructure of nontrivial size \cite{RuegLak16} \cite{RuegLakLatt18}   \cite{Merkel11} and in chiral solids \cite{CossChiral82}.  Chirality was observed to give rise to stretch - twist or squeeze twist coupling in  rib lattices \cite{HaLakChiral16} \cite{ReaLakChir19}  \cite{HaLakChiral20} and in surface lattices such as the gyroid \cite{ReasaLakChiral20}.  Stretch-twist coupling was observed in bone \cite{LakesBone81} and in fiber bundles of tendon \cite{BuchananVanderTendon17}. Chiral cholesteric elastomers \cite{CossChiralPRL09} were predicted to twist in response to stretching with a characteristic length predicted to be on the order of 10 nm. 
\par
Chirality is shown in this study to have an effect on thermal conductivity through a passive solid. Flow in chiral media can give rise to circulating flow via a third rank tensor. The concept is applicable to flow of heat, fluids in porous media, or electric current. The third rank contribution due to chirality is distinct from odd effects associated with an asymmetric second rank property tensor due to time reversal asymmetry that can arise from an external supply of energy   or from dissipation \cite{Pipkin63}.    
\par

\section{Analysis} \label{Analysis}
Consider flow of heat through a solid that may be chiral. 
Suppose the current $J$ is given in terms of a field via a superposition of contributions from a second rank and a third rank tensor. 
\par
The current is $J_{i} = g_{ij} E_{j} + k_{ijk} E_{j,k}$ with $E_{j}$ as the temperature gradient field,  $g_{ij}$ as the classical thermal conductivity and $k_{ijk}$ as a third rank material property tensor.
 An inverse form is 
\begin{equation} \label{eq:thirdrank}
 E_{i} = \rho_{ij} J_{j} + \rho_{ijk} J_{j,k}
\end{equation} 
in which $\rho_{ij}$ is the classical thermal resistivity, and $\rho_{ijk}$ is a material property tensor of third rank.  The usual Einstein summation convention for repeated indices is used. 
\par
Suppose the material is isotropic. 
There is one isotropic third rank tensor, the permutation symbol $e_{ijk}$.  It is antisymmetric: $e_{ijk} = -e_{ikj}$. The gradient introduces dimensions of inverse length. Define a characteristic length $\ell_{th}$ with dimensions of length such that $\rho_{T} = \frac{R}{\ell_{th}}\rho_{L}$. The resistivity relation becomes, for a round specimen of radius $R$, 
\begin{equation} \label{eq:thirdrankiso}
 E_{i} =   (\rho_{L} \delta_{ij} J_{j} + \rho_{T}R e_{ijk} J_{j,k})
\end{equation} 
in which $\rho_{L}$ and $\rho_{T}$ are the longitudinal and transverse resistivity coefficients respectively. 
\par
Recognizing $curl J = e_{ijk} J_{j,k}$ and using Green's theorem, then at the periphery at outer radius $R$, $E_{L} = \rho_{L}J_{L}  + 2 \rho_{T}J_{T}$. 
\par
At first sight one might think the second term vanishes in Equation  \ref{eq:thirdrankiso} in view of the fact that curl of a gradient equals zero.  However the point of this research is that the derivatives of the current are not subsumed in a gradient. The derivative $J_{j,k}$ need not equal $J_{k,j}$.  If there is chirality, then a circulating current can occur, not just a gradient.  In any case, these considerations only apply for materials that are isotropic with respect to all physical properties. The gyroid is structurally cubic, though classical elastic properties are approximately isotropic. 
 \par
The transverse circulating current is
\begin{equation} \label{eq:thirdrankiso2}
J_{T} = (E_{L}  -  \rho_{L}J_{L} ) \frac{\ell_{th}} {R} \frac{1} {2\rho_{L}} 
\end{equation}
so the nonclassical flow is proportional to a characteristic length as with phenomena in nonclassical elasticity and piezoelectricity. 
\par
\begin{figure}  [!htb]
		\includegraphics[width=0.39 \textwidth]{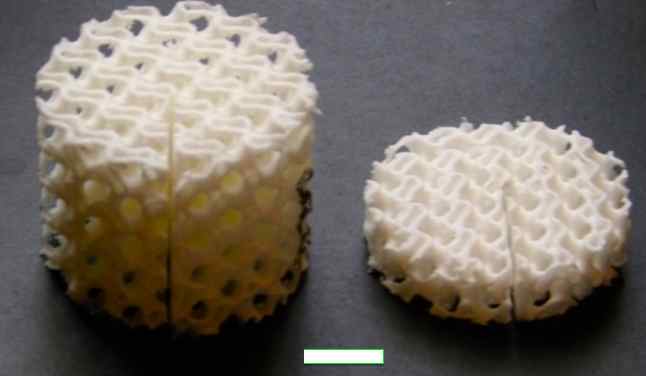}
		\caption{\label{gyroidpic} Chiral gyroid lattice. Cylinder and disc with cut slots facing forward. Scale bar: 1 cm. }
 \end{figure}
 
\section{Experiment} \label{Experiment}
To experimentally study heat flow in a chiral solid, chiral gyroid surface lattices were made of nylon polymer via 3D printing. The lattices \cite{ReasaLakChiral20} had a surface wall thickness of 0.4 mm and a cell size of 6 mm. The solid material was a nylon polymer PA 2200 for which the density was $\rho_{s}$ = 0.98 g/cm$^{3}$. The ratio of lattice density to the solid density was $\rho/\rho_{s}$ =  0.195 for the chiral gyroid.   Gyroid lattices were originally developed  \cite{Schoen68, Schoen12}  for use in stiff, strong and lightweight structures. 
An infrared thermal camera (FLIR TG 165) was used to study patterns of heat flow. Experiments were repeated and multiple images were captured.   
\par 

\begin{figure}  [!htb]
		\includegraphics[width=0.15\textwidth]{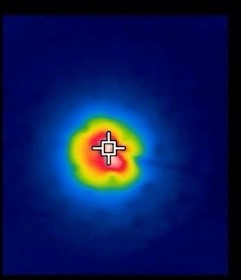}
		\includegraphics[width=0.15\textwidth]{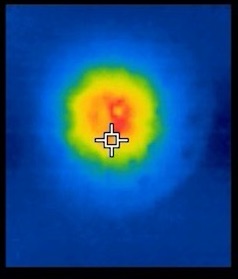}
		\includegraphics[width=0.15\textwidth]{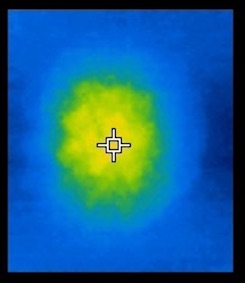}
		\includegraphics[width=0.15\textwidth]{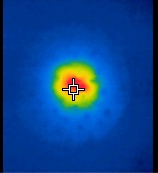}
		\includegraphics[width=0.15\textwidth]{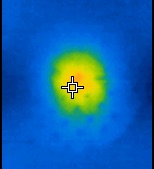}
		\includegraphics[width=0.15\textwidth]{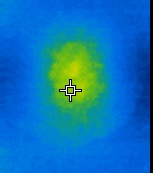}
		\caption{\label{discgyroid} Chiral gyroid lattice. Thermal images, circular disc with slot at the right. Heat source is at the center.\\
		Top series, left, time 30 sec, center, 90 sec, right, 180 sec.\\
		Bottom series, left, time 30 sec, center, 60 sec, right, 120 sec.  }
\end{figure}
To probe possible asymmetry of the second rank thermal resistivity tensor $\rho_{ij}$ in Equation  \ref{eq:thirdrank},   a disc with a radial slot was heated at the center and temperature distribution observed.   The method is after Soret \cite{Soret1893} \cite{Soret1894}. The heat flow due to the diagonal components of $\rho_{ij}$ is radial.  
An asymmetric second rank resistivity tensor with off diagonal terms (Equation  \ref{eq:resistivitymatrix}) generates a transverse flow. 
\begin{equation}  \label{eq:resistivitymatrix}
\rho_{ij} = 
\left( \begin{array}{ccc} 
\rho_{1} & \rho_{g}  & 0  \\ 
-\rho_{g} & \rho_{1} & 0 \\ 
0 & 0 & \rho_{1} 
\end{array} \right) 
\end{equation}
If there is a circulating flux due to an asymmetric conductivity tensor, then the slot will interrupt the circumferential component of the flow and there will be build up of heat along one edge. The method was tried with a single crystal of gypsum \cite{Soret1893} \cite{Soret1894} but no evidence of circulating flow was found.  
\par
  To probe such circulation, a gyroid disc  8 mm thick and 35 mm in diameter of gyroid lattice was  cut and a slot was saw cut from periphery to center (Figure \ref{gyroidpic}).   The disc was heated at its center with a soldering iron adjusted to minimum heat. Heating was applied for 60 sec and infrared thermal images were taken after time delays of 30, 60 and 120 sec. Figure \ref{discgyroid} shows thermal images of the gyroid disc specimen. The slot on the right is not apparent and there is no evidence of heat build up at the slot edges  in these experiments or in repeated tests with different heat input times   as shown in the second series of images at the bottom.   There was therefore no evidence of circulating flux in the chiral disc, hence no evidence of an asymmetric second rank thermal conduction tensor. Moreover, the circular shape of the expanding heated region indicates directional isotropy of the thermal conductivity or resistivity. By contrast, heating of wood by the same method resulted in an elongated elliptic shape of the warm region revealing the expected directional anisotropy. The gyroid is structurally cubic and its modulus and thermal conductivity are isotropic. An asymmetric conductivity tensor requires a direction that could be provided by anisotropy or by an external stimulus as in the case of the Faraday effect. Chirality does not suffice to give rise to off diagonal second rank terms because chirality   was shown \cite{Nye} to have   no effect on second rank tensor properties. 
\par
\begin{figure}  [!htb]
		\includegraphics[width=0.15 \textwidth]{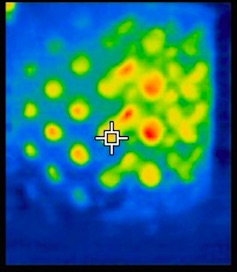}
		\includegraphics[width=0.15 \textwidth]{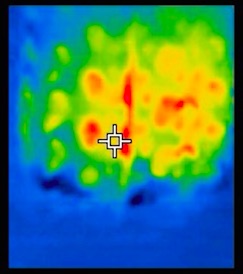}
		\includegraphics[width=0.15 \textwidth]{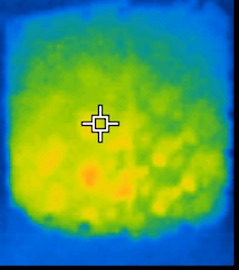}
		\caption{\label{cylgyroid} Chiral gyroid lattice. Thermal images:  Left, intact cylinder.    Center, cylinder with slot at the front. Right, cylinder with slot at the front, time 60 sec longer.  Heat source was at the bottom. }
 \end{figure}

Cylindrical gyroid specimens 35 mm in diameter and 30 mm long were prepared to explore, in three dimensions, circulating thermal flux resulting from a longitudinal thermal gradient.
   Lattice specimens were heated from below with a hot plate. An oblique pattern of temperature variations was seen in a chiral cylinder but not in a non-chiral cylinder. This suggests spiral flow in the chiral lattice.  To explicitly discern the possibility of spiral flow,    one chiral gyroid cylinder was prepared with a longitudinal slot saw cut from the periphery to near the center (Figure \ref{gyroidpic}). A second chiral cylinder was intact. Both were heated from one end using a hot plate (52$^{\circ}$C) for 60 sec, then removed from the heat source     and placed on an insulating surface (Figure \ref{gyroidpic}).  Infrared images were taken 60 sec after removal.  
Figure \ref{cylgyroid} shows   representative   thermal images of the cylinder specimens with and without a slot.     The image on the right is for a slotted cylinder at a time 60 sec longer.  
For cylinders, infrared images    in these experiments    (Figure \ref{cylgyroid})  and repeated tests   reveal a build up of heat along one edge of the slot, hence a circulating component of flux perpendicular to the predominant heat flow along the    vertical  axis of the cylinder. Consequently the second term in Eq. \ref{eq:thirdrankiso}, hence the characteristic length, is nonzero. 

The methods suffice to reveal the presence of an effect of an asymmetric component to $\rho_{ij}$ and of a third rank contribution $\rho_{ijk}$. A quantitative measure of thermal resistivity $\rho_{L}$ can be obtained via conventional methods. To determine the nonclassical characteristic length $\ell_{th}$, one can prepare an array of cylinders with oblique slots  at different angles. The angle corresponding to zero heat build up reveals the pitch of the spiral heat flow, hence the ratio of the resistivity terms in Equation (\ref{eq:thirdrankiso}).   Similarly, oblique slots in discs can reveal the ratio of asymmetric second rank terms to symmetric terms.  
\par
  Characteristic lengths in generalized elasticity, in piezoelectricity, and in thermal flow will be related to physical length scales in the material. Characteristic length scales have long been known \cite{BursianTrunov74} in piezoelectric materials. There is a fourth rank sensitivity term in addition to the usual third rank sensitivity. Materials with such gradient effects are now called flexoelectric \cite{Zubko2021}.  If the dominant length scale is the spacing between atoms, a macroscopic scale experiment will not suffice to reveal effects;  a nano-scale experiment will be needed. Indeed, the early macroscopic   thermal  experiments  \cite{Soret1893}  \cite{Soret1894} would not detect effects, if present, on the atomic scale in single crystals. In any case, second rank effects are not expected unless the material does not obey time reversal symmetry as in viscoelastic materials or in active materials. The third rank properties can occur in passive materials and are observable provided material length scales are not too small compared with experimental length scales. 
\par

\section{Discussion} \label{Discussion}
In active materials, internal degrees of freedom are excited by an external power source; time reversal symmetry is broken. An antisymmetric diffusivity, denoted ``odd'', emerges in a chiral random walk model \cite{HargusPRL21} in an active medium with an external supply of energy. The  diffusivity tensor for such materials contains both a symmetric and an antisymmetric component.  Molecular dynamics simulations of chiral active matter indicate such effects can occur. Since there is no rank-two tensor in three dimensions which is both isotropic and antisymmetric, attention was given to two-dimensional diffusion. In a related study, shear and ``odd''  viscosity values  were inferred in a model system consisting of actively torqued dumbbells using molecular dynamics simulations \cite{HargusChemPhys20}. 
\par
Time reversal asymmetry also occurs in passive viscoelastic materials. Wood, which is viscoelastic and has orthotropic anisotropy, is observed to exhibit  \cite{Neuhaus1983} \cite{Hering2012}  a compliance which is asymmetric with respect to exchange of pairs of indices   as anticipated theoretically \cite{Pipkin63}.   It was not, however, called  ``odd''.  
\par
Passive materials were considered more recently in the context of odd properties \cite{Lier2022} and analyzed to show that such effects can occur passive chiral viscoelastic fluids. Odd properties were analyzed for hypothetical active Cosserat solids \cite{Surowka2023}. Wave modes were found to depend on both odd and Cosserat aspects.   
\par
As for other second rank tensor properties, consider thermo-elasticity in chiral materials. The second rank thermal expansion tensor is proportional to the strain, which by definition is symmetric. Therefore the thermal expansion tensor cannot be asymmetric. Nevertheless, effects of chirality occur in the thermal response provided the material has a characteristic length scale.  A chiral Cosserat analysis \cite{IesanCossTherm10} incorporating rotational degrees of freedom includes a temperature dependent term that contributes to the couple stress or moment per area, hence to the overall deformation. The analysis predicts that a temperature change will give rise to a twisting effect in a chiral thermo-elastic rod. As with squeeze-twist coupling in elastic solids, heat-twist coupling effects cannot be understood within classical elasticity. The thermal twist was observed experimentally \cite{Hestetune2022} in chiral gyroid lattices and interpreted via Cosserat elasticity which has a characteristic length scale.  
\par
Electrical conductivity is represented by a second rank tensor. The second rank dielectric permittivity tensor can be asymmetric in the presence of a magnetic field, giving rise to the Faraday effect as is well known. The magnetic field introduces anisotropy and prevents time reversal symmetry.  
If one introduces in electrical conductivity  a third rank term associated with chirality, then the resulting circulating electric current due to an axial applied voltage will give rise to an axial magnetic field, which is detectable. 
\par
Asymmetry in second rank properties requires anisotropy  \cite{Avron1998}  and violation of time reversal invariance  \cite{Onsager1, Onsager2}. Active materials represent one way to obtain time reversal non-invariance; viscoelastic materials represent another way. 
\par
Chirality in passive materials does not give rise to asymmetric second rank properties but does give rise to third rank properties, even if the material is isotropic. The thermal conductivity third rank tensor differs from the piezoelectric third rank tensor in that the thermal tensor can be antisymmetric. By contrast, classical piezoelectricity requires $d_{ijk} = d_{ikj}$ because the stress is symmetric in classical elasticity. In piezoelectricity the 27 independent coefficients are reduced to 18, allowing representation as a 3 by 6 matrix \cite{Nye} but this cannot be done in the thermal conductivity case. Also there is no characteristic length scale in classical piezoelectricity. 
\par
\section{Conclusion} \label{Conclusion}
To conclude, third rank thermal properties are observed and a method is provided for measuring them in comparison with the usual second rank properties.   No evidence of asymmetric second rank properties was observed    in these solids.

\end{document}